\documentclass[aps,prl,groupedaddress]{revtex4-2}

\usepackage{amsmath,amssymb,bm}
\usepackage[colorlinks=true,linkcolor=blue,citecolor=blue,urlcolor=blue]{hyperref}

\begin{document}

\title{Contextuality as an External Bookkeeping Cost under Fixed Shared-State Semantics}

\author{Song-Ju Kim}
\email{kim@sobin.org}
\affiliation{SOBIN Institute LLC, Kawanishi, Hyogo, Japan}

\date{\today}

\begin{abstract}
Contextuality is a central feature distinguishing quantum from classical probability theories, but its operational meaning is often stated only qualitatively.
In this Letter, we study a simple information-theoretic question: how much additional contextual information must a classical simulation introduce when it tries to keep a shared internal description fixed across contexts?
To make this question precise, we analyze a minimal external-label simulation model in which the remaining context dependence is carried only by an auxiliary label.
For this model, we define an obstruction cost as the minimum mutual information between the context and the auxiliary label required to reproduce the observed statistics.
We then prove a conservative quantitative lower bound: any linear witness that separates the observed statistics from the zero-obstruction set yields a positive lower bound on this cost.
We do not claim that this bound is tight, and we do not claim that the simulation model covers every possible classical architecture.
Its role is narrower and more explicit: under fixed shared-state semantics, contextuality can be read as a certificate of irreducible external bookkeeping cost in a simple and well-defined simulation model.
\end{abstract}

\maketitle

\section{Introduction}

A central theme in the foundations of quantum theory is to clarify \emph{why} quantum probability departs from classical probability, beyond axiomatic postulates.
One prominent structural feature is \emph{contextuality}: the impossibility of representing the outcome statistics of multiple operational contexts as marginals of a single joint distribution over context-independent latent variables
\cite{KochenSpecker1967,KCBS,Spekkens2005,AbramskyBrandenburger2011}.
Contextuality is often presented as a nonclassical anomaly.
At the same time, quantitative and resource-oriented approaches have shown that it can also be meaningful to ask how strongly contextual a given scenario is, or what task-relevant role contextuality may play
\cite{AbramskyBarbosaMansfield2017,Howard2014}.
In the present work, we ask a different but related question: what explicit representational overhead appears when one tries to keep a shared internal description fixed across contexts?

Motivated by recent work on the Quantum Tug-of-War (QTOW) framework~\cite{QTOW},
we consider operational scenarios in which an agent or device is constrained to maintain a \emph{single internal state} whose meaning remains fixed across contexts, while context dependence is realized through context-dependent operations.
QTOW shows that, for certain collections of contexts, enforcing such shared-state semantics leads to contextual statistics that are incompatible with a noncontextual classical embedding.
The goal of the present Letter is not to extend QTOW, and it is not to make claims about performance or learning efficiency.
Our aim is narrower: we re-express the same obstruction in an information-theoretic language that makes one concrete bookkeeping cost explicit.

Our guiding question is the following:
\begin{quote}
\emph{If a contextual single-state operational model is given, how much additional context information must a classical simulation carry when it tries to keep the shared internal description fixed?}
\end{quote}

We do not attempt to answer this question for every possible classical architecture.
Instead, we analyze one simple and explicit simulation class in which the remaining context dependence is carried by an auxiliary label.
Within this class, we define an obstruction cost and prove a witness-based lower bound on it.
This gives a limited but precise statement: in this simulation model, a witness violation above the zero-obstruction threshold certifies a nonzero external contextual bookkeeping cost.

We use ``single-state semantics'' to formalize compressed representations in which the internal state is not allowed to split into context-indexed meaning, as in minimal-memory devices or models that restrict internal registers.

\section{Single-State Constraints and Contextuality}

We consider an operational model specified by a set of contexts $c \in \mathcal{C}$, an internal state $S$, and outcomes $y \in \mathcal{Y}$.
Operationally, a context corresponds to an intervention that produces outcomes with conditional distribution $P(y|c)$, or more generally $P(y|c,s)$ when conditioned on the internal state.

The \emph{single-state semantic requirement} demands that the internal state retain the same meaning across all contexts.
Information-theoretically, this is expressed as
\begin{equation}
I(S;C)=0,
\label{eq:single_state_semantics}
\end{equation}
where $C$ denotes the random variable over contexts and $I(\cdot;\cdot)$ is the mutual information.
Equation~(\ref{eq:single_state_semantics}) formalizes the absence of explicit contextual labeling in the internal representation and is purely representational, making no assumptions about dynamics.
It is not a definition of noncontextuality.
Rather, it is a modeling constraint that can be imposed even in situations where the observed statistics are operationally contextual.

In QTOW, this constraint is enforced structurally: the same underlying state is maintained while different contexts act through different operations.
For certain collections of contexts, however, the resulting statistics cannot be reproduced by any classical probabilistic model that preserves Eq.~(\ref{eq:single_state_semantics}).
This incompatibility is one operational manifestation of contextuality
\cite{Spekkens2005,AbramskyBrandenburger2011}.

When a shared-state description of the form in Eq.~(\ref{eq:single_state_semantics}) is not available, there are at least two generic responses.
One is to let the internal representation itself absorb contextual dependence, which can be written schematically as
\begin{equation}
I(S;C) > 0.
\label{eq:internal_leakage}
\end{equation}
The other is to keep the shared internal description fixed and move the remaining context dependence to an auxiliary external variable.
In the present Letter we analyze the second route, because in that case one can define an explicit obstruction cost and derive a quantitative lower bound on it.
We do not derive a general lower bound on the internal leakage quantity in Eq.~(\ref{eq:internal_leakage}) from observed statistics.

\section{A Minimal Obstruction Measure and a Witness Lower Bound}

We now make the external bookkeeping question precise in a minimal classical simulation model.
Here $M$ is an abstract auxiliary label that specifies which context-independent response rule is used in the simulation.
It should be understood as a bookkeeping variable, not as a claim about a unique physical memory register, a learned hidden state, or a full agent architecture.
Accordingly, Eq.~(\ref{eq:rule_mixture_model}) should be read as a one-shot classical simulation model whose only task is to reproduce the conditional statistics $P(y|c)$.
We use this restricted class because it is the simplest setting in which an explicit external contextual cost can be defined and bounded.

\begin{equation}
P(y|c)=\sum_{m\in\mathcal{M}} \eta(m|c)\,\tau(y|m).
\label{eq:rule_mixture_model}
\end{equation}
Here $\eta(m|c)$ is a context-dependent distribution over auxiliary labels, and $\tau(y|m)$ is a context-independent response law.
We do not claim that Eq.~(\ref{eq:rule_mixture_model}) covers every possible classical simulation architecture.
This factorization is a structural restriction on the simulation class.
It assumes that the remaining context dependence is carried entirely by the auxiliary label $M$, so that $Y \perp C \mid M$.
Our result is proved only within this explicit external-label simulation model.

Fix a full-support reference distribution $\pi(c)>0$ on $\mathcal{C}$.
For a representation of the form in Eq.~(\ref{eq:rule_mixture_model}), let
\begin{equation}
p_{\pi}(c,m):=\pi(c)\eta(m|c),
\end{equation}
and let $I_{\pi}(C;M)$ denote the mutual information computed from $p_{\pi}(c,m)$.
Throughout this section, logarithms are taken in base 2.

We define the obstruction cost of $P$ by
\begin{equation}
\mathcal{O}_{\pi}(P)
:=
\inf_{\eta,\tau}
I_{\pi}(C;M),
\qquad
\text{subject to Eq.~(\ref{eq:rule_mixture_model}) reproducing } P(y|c).
\label{eq:obstruction_cost}
\end{equation}
Thus, $\mathcal{O}_{\pi}(P)$ is the minimum amount of context information that must be carried by the auxiliary label in this simulation class.

We also define the $\pi$-weighted total variation distance between two conditional models $P$ and $Q$ by
\begin{equation}
d_{\pi}(P,Q)
:=
\frac{1}{2}\sum_{c\in\mathcal{C}}\pi(c)\,
\bigl\|P(\cdot|c)-Q(\cdot|c)\bigr\|_{1}.
\label{eq:weighted_tv}
\end{equation}

Let
\begin{equation}
\mathcal{N}_{0}
:=
\{Q:\mathcal{O}_{\pi}(Q)=0\}
\label{eq:zero_obstruction_set}
\end{equation}
be the zero-obstruction set.
In the present simulation model, $Q\in\mathcal{N}_{0}$ means that the simulation can be carried out with an auxiliary label independent of the context.

For a linear functional
\begin{equation}
W(P)=\sum_{c\in\mathcal{C}}\sum_{y\in\mathcal{Y}} w_{c,y}\,P(y|c),
\label{eq:linear_witness}
\end{equation}
define the corresponding zero-obstruction threshold
\begin{equation}
\beta_{0}:=\sup_{Q\in\mathcal{N}_{0}} W(Q),
\label{eq:beta0}
\end{equation}
and the witness violation
\begin{equation}
\Delta_{W}(P):=\max\{0,\;W(P)-\beta_{0}\}.
\label{eq:witness_violation}
\end{equation}
Here a witness means a linear functional on the observed conditional statistics.
Its role is to separate the zero-obstruction set $\mathcal N_0$ from statistics that lie outside it:
if $W(P)>\beta_0:=\sup_{Q\in\mathcal N_0}W(Q)$, then $P\notin\mathcal N_0$.
We also write
\begin{equation}
L_{W,\pi}
:=
\sup_{P\neq Q}
\frac{|W(P)-W(Q)|}{d_{\pi}(P,Q)}
\label{eq:lipschitz_constant}
\end{equation}
for the Lipschitz constant of $W$ with respect to $d_{\pi}$.
This constant measures how strongly the witness value can change under a given change in the conditional statistics, as measured by the weighted total variation distance $d_{\pi}$.

This theorem should not be read as a full characterization of contextuality in general operational theories.
Rather, it quantifies the obstruction cost within the restricted external-label simulation model of Eq.~(\ref{eq:rule_mixture_model}), which is introduced here as a minimal bookkeeping model motivated by fixed shared-state semantics.
\vspace{5mm}

\paragraph{Theorem 1.}
For every conditional statistics $P(y|c)$ admitting a representation of the form in Eq.~(\ref{eq:rule_mixture_model}),
\begin{equation}
\mathcal{O}_{\pi}(P)
\;\ge\;
\frac{2}{\ln 2}\,
\frac{\Delta_{W}(P)^{2}}{L_{W,\pi}^{2}}.
\label{eq:main_lower_bound}
\end{equation}

\paragraph{Proof.}
Take any representation of $P$ of the form
\begin{equation}
P(y|c)=\sum_{m}\eta(m|c)\tau(y|m).
\end{equation}
Let
\begin{equation}
p_{\pi}(c,m)=\pi(c)\eta(m|c),
\qquad
p_{\pi}(m)=\sum_{c}\pi(c)\eta(m|c).
\end{equation}
From this representation, define
\begin{equation}
Q(y|c):=\sum_{m} p_{\pi}(m)\tau(y|m).
\label{eq:q_def}
\end{equation}
Since the right-hand side does not depend on $c$, we have $\mathcal{O}_{\pi}(Q)=0$, hence $Q\in\mathcal{N}_{0}$.

Now define the joint distributions
\begin{equation}
p_{CY}(c,y):=\pi(c)P(y|c),
\qquad
q_{CY}(c,y):=\pi(c)Q(y|c).
\end{equation}
The distribution $q_{CY}$ is obtained from the product distribution
\begin{equation}
\widetilde{p}_{CM}(c,m):=\pi(c)p_{\pi}(m)
\end{equation}
by passing $(c,m)$ through the same channel $\tau(y|m)$.
Therefore, by the data-processing inequality \cite{CoverThomas},
\begin{equation}
D_{\mathrm{KL}}(p_{CY}\|q_{CY})
\le
D_{\mathrm{KL}}(p_{\pi}(c,m)\|\widetilde{p}_{CM}(c,m))
=
I_{\pi}(C;M).
\label{eq:dpi_step}
\end{equation}
By Pinsker's inequality \cite{CoverThomas},
\begin{equation}
d_{\pi}(P,Q)^{2}
=
\mathrm{TV}(p_{CY},q_{CY})^{2}
\le
\frac{\ln 2}{2}\,
D_{\mathrm{KL}}(p_{CY}\|q_{CY}).
\end{equation}
Combining this with Eq.~(\ref{eq:dpi_step}) gives
\begin{equation}
d_{\pi}(P,Q)^{2}
\le
\frac{\ln 2}{2}\,
I_{\pi}(C;M).
\label{eq:pinsker_step}
\end{equation}

On the other hand, since $Q\in\mathcal{N}_{0}$, we have $W(Q)\le \beta_{0}$.
Hence
\begin{equation}
\Delta_{W}(P)
\le
W(P)-W(Q)
\le
|W(P)-W(Q)|
\le
L_{W,\pi}\, d_{\pi}(P,Q).
\label{eq:witness_step}
\end{equation}
Combining Eqs.~(\ref{eq:pinsker_step}) and (\ref{eq:witness_step}) yields
\begin{equation}
\Delta_{W}(P)^{2}
\le
L_{W,\pi}^{2}\,\frac{\ln 2}{2}\,I_{\pi}(C;M).
\end{equation}
Therefore,
\begin{equation}
I_{\pi}(C;M)
\ge
\frac{2}{\ln 2}\,
\frac{\Delta_{W}(P)^{2}}{L_{W,\pi}^{2}}.
\end{equation}
Since this is true for every representation of $P$ of the form in Eq.~(\ref{eq:rule_mixture_model}), taking the infimum over all such representations proves Eq.~(\ref{eq:main_lower_bound}).
\hfill$\square$

\paragraph{Corollary 1.}
Let $W_{\ast}$ be a chosen linear witness for the operational scenario of interest, and let
\begin{equation}
\beta_{\ast}:=\sup_{Q\in\mathcal{N}_{0}} W_{\ast}(Q).
\end{equation}
If
\begin{equation}
W_{\ast}(P)>\beta_{\ast},
\end{equation}
then
\begin{equation}
\mathcal{O}_{\pi}(P)
\;\ge\;
\frac{2}{\ln 2}\,
\frac{\bigl(W_{\ast}(P)-\beta_{\ast}\bigr)^{2}}{L_{W_{\ast},\pi}^{2}}
\;>\;0.
\label{eq:chosen_witness_corollary}
\end{equation}
Thus, once a concrete witness is fixed, any violation above the zero-obstruction threshold certifies a nonzero external contextual bookkeeping cost in the simulation class of Eq.~(\ref{eq:rule_mixture_model}).

A simple explicit upper bound on $L_{W,\pi}$ is also available.
If
\begin{equation}
W(P)=\sum_{c,y} w_{c,y}P(y|c),
\end{equation}
then
\begin{equation}
L_{W,\pi}
\le
2\max_{c\in\mathcal{C}}
\frac{\max_{y\in\mathcal{Y}}|w_{c,y}|}{\pi(c)}.
\label{eq:lipschitz_simple_bound}
\end{equation}
In the uniform case $\pi(c)=1/|\mathcal{C}|$, this becomes
\begin{equation}
L_{W,\pi}
\le
2|\mathcal{C}|\max_{c,y}|w_{c,y}|.
\end{equation}

The bound in Theorem~1 is conservative.
We do not claim that it is tight, and we do not claim that Eq.~(\ref{eq:rule_mixture_model}) exhausts all possible classical architectures.
Its role is narrower: it gives an explicit lower bound, within a simple and well-defined simulation model, on how much contextual information must be carried externally when one insists on fixed shared-state semantics.

\section{Discussion}

The result derived here does not show that quantum theory is the only possible description of single-state operational scenarios.
What it does show is narrower and more concrete.
In the minimal simulation model of Sec.~III, contextual dependence cannot be removed for free once one insists on a fixed shared-state description.
If the simulation keeps the shared internal description fixed and allows the remaining dependence to appear only through an auxiliary label, then a witness violation above the zero-obstruction threshold forces a positive mutual-information cost between the context and that label.

We therefore do not claim a universal contextuality monotone, and we do not claim a full no-go theorem for every classical architecture.
Our point is more modest: the model of Eq.~(\ref{eq:rule_mixture_model}) gives one honest and mathematically explicit sense in which contextuality creates a classical bookkeeping overhead.
In this sense, contextuality functions here as a witness of irreducible representational cost in a simple external-label model, rather than only as a binary nonclassical anomaly.

This perspective also helps explain why state-operation descriptions of the kind used in quantum theory can be natural in scenarios with fixed shared-state semantics.
Such descriptions allow different contexts to act through different operations without requiring an explicit external context label in the same way as the minimal classical simulation considered here.
We do not claim that the present argument alone derives quantum theory.
Rather, it identifies a limited representational pressure that points away from naive shared-state classical descriptions when contextual statistics are present.

Several extensions remain open, including sharper evaluation of $\mathcal{O}_{\pi}(P)$ for specific operational constructions, approximate simulations via rate--distortion trade-offs, and connections to quantitative and resource-theoretic approaches to contextuality
\cite{AbramskyBarbosaMansfield2017,Howard2014}.
Extending the present witness-based lower bound beyond the minimal simulation class of Eq.~(\ref{eq:rule_mixture_model}), and relating it directly to internal leakage quantities such as $I(S;C)$, are natural next steps.
These directions may sharpen the present bound while leaving the basic conceptual point unchanged.

\section*{Declaration of competing interest}
The author declares no competing interests.

\vspace{5mm}

\section*{Acknowledgments}

The author used ChatGPT (OpenAI) for English editing and takes full responsibility for the final version.


\end{document}